\documentclass[12pt]{article}
\usepackage{times,amssymb,euscript,epsfig,latexsym,amsmath}
\usepackage[normalem]{ulem}
\usepackage{epsfig,graphicx,subfig}
\usepackage{color}
\begin{document}
\newcommand{\todo}[1]{\vspace{5 mm}\par \noindent
\framebox{\begin{minipage}[c]{0.95 \textwidth}
\tt #1 \end{minipage}}\vspace{5 mm}\par}
\newtheorem{thm}{Theorem}
\newtheorem{prop}[thm]{Proposition}
\date{}
\title{ \Large{Motion of a Buoyant Vortex Patch}}
\author{Banavara N. Shashikanth\footnote{Mechanical and Aerospace Engineering
Department, MSC 3450, PO Box 30001, New Mexico State University,
Las Cruces, NM 88003, USA. E-mail:shashi@nmsu.edu} and Rangachari Kidambi \footnote{Computational \& Theoretical Fluid Dynamics Division, National Aerospace Laboratories, Bengaluru 560017, India. E-mail : kidambi@nal.res.in}}

\maketitle

\abstract{The motion of a two-dimensional buoyant vortex patch, i.e. a vortex patch with a uniform density different from the uniform density of the surrounding fluid, is analysed in terms of evolution equations for the motion of its centroid, deformation of its boundary and the strength of a vortex sheet which is essential to enforce pressure continuity across the boundary. The equations for the centroid are derived by a linear momentum analysis and that for the sheet strength by applying Euler's equations on the  boundary, while the boundary deformation is studied in the centroid-fixed frame. A complicated coupled set of equations is obtained which, to the best of our knowledge, has not been derived before. The evolution of the sheet strength is obtained as an integral equation. The equations are also examined in the limit of a patch of vanishing size or a buoyant point vortex. }
 
\section{Introduction}

A vortex patch is a finite region of constant vorticity, surrounded by an irrotational flow. In the context of inviscid fluid flows, many studies have been published devoted to the dynamics of vortex patches in fluids of a single uniform density, henceforth also referred to as neutrally buoyant vortex patches. The Rankine and Kirchhoff vortex patches \cite{lamb} are the simplest and best-known examples denoting a circular patch and an elliptical patch, respectively, rotating in fluid at rest at infinity.  Deem and Zabusky in a pioneering paper \cite{DeZa1979}  transformed the evolution equation of a patch to an evolution equation for its boundary alone. Based on this approach, named contour dynamics,  they presented numerical evidence of a class of rotating isolated patches of more general shapes and a class of translating patch pairs of opposite-signed vorticity; see also \cite{ZaHuRo1979}. The rotating solutions, termed as $V$-states,  have $m$-fold symmetry ($m$ integer), i.e. $m$ axes of symmetry. The Kirchhoff patch corresponds to $m=2$.   Burbea \cite{burb} used analytical techniques to derive some of the $V$-states. Moore and Saffman \cite{MoSa1971}---and later Kida \cite{Ki1981}---studied, as a first level approximation, the effect of a uniform shear flow, i.e. a flow in which the velocity field is linear in the coordinates, on the dynamics of an isolated elliptic patch. The patch continues to retain its elliptic form but its aspect ratio and angular velocity get modified. A  Hamiltonian generalization of this is the $N$ patch moment model of Melander, Zabusky and Styczek \cite{MeZaSt1986}.  More investigations about the opposite-signed translating pair were made by Pierrehumbert \cite{Pi1980} and Yang and Kubota \cite{YaKu1994}. Saffman and Szeto \cite {SaSz1980} investigated same-signed patch pairs that rotate about each other. Linear and rotating arrays of vortex patches have also been studied; see, for example, \cite{SaSz1981, PiWi1981, crowd}.  Dritschel \cite{Dr1988} has used contour dynamics and its improvements like contour surgery to study different features of patch evolution such as filamentation and merging. Patches have also been investigated in a Hamiltonian framework. Marsden and Weinstein \cite{MaWe1983}, among other things, placed the Hamiltonian structure of singular vortex models, such as point vortices and vortex filaments, as also that of a vortex patch, in the wider framework of modern theories of Hamiltonian systems with symmetry and reduction. Exploiting this Hamiltonian structure, the nonlinear stability of patches was analyzed in \cite{WaPu1985, Wa1986, Wa1988}.

On the other hand, the motion of vortex patches, with density variations present, even though very relevant to geophysical and atmospheric flows, has attracted relatively less attention. Arendt \cite{Ar1993a, Ar1993b} studied such patches as a model of sunspots while \cite{Ar1996} presented point vortex models of vortex motion in a stratified fluid. Steady stratification and an essentially hydrostatic pressure field were assumed.  An important finding of Arendt's work is that such vortices, under the influence of the stratification, have a self-propulsion transverse to gravity and that this propulsion is logarithmically singular in the ratio of the cross-section to the density scale height. In an earlier paper,  Saffman \cite{Sa1972} developed a model for a buoyant point vortex pair in a stratified atmosphere and concluded that they performed oscillatory motions in a vertical direction, the distance between them remaining constant. Hill \cite{hill1975} numerically confirmed Saffman's predictions for small times.  High frequency oscillations were found to be possible in the  motion of a single, buoyant rectilinear vortex, of small core cross-section and that neglecting the finite cross section except in the buoyancy term, leads to the point vortex equations under the action of a force \cite{Sa1992}.  Both Arendt's and Saffman's papers also contain references to some earlier work on the topic. 

     Modeling vortex interactions in the presence of a continuously varying density field is, in general, a very difficult task; especially obtaining analogues of some well-studied low-dimensional models in homogeneous flows, such as an isolated vortex patch or the classical  $N$-point-vortex model. A  relatively simpler problem is where the density field is modeled as a step function. Specifically, the uniform density of the patch $\rho_i$ is assumed to be different from the uniform density $\rho_o$ of the fluid external to it. A popular assumption in the literature on surface-tension-free interfaces has been the use of pressure continuity at the interface. For patches with density jumps, Euler's equations implies a jump in the tangential velocity at the boundary and consequently a vortex sheet. The resulting models, involving vortex patches bounded by vortex sheets, have been referred to in the literature as Prandtl-Batchelor flows or sheet patches \cite{Sa1992}; we use the term bouyant vortex patch (BVP) in this paper. \textcolor{blue}{ Note that, in these models, vorticity generation happens only at the patch boundary, due to baroclinicity; elsewhere, the fluid is barotropic. } Sheet patch studies have mostly involved computation of equilibria in a variety of background flows (\cite{frei}, \cite{kao}, \cite{zan}). For example, in the important paper of Kao \& Caflisch \cite{kao} invariant translating vortex patch shapes as a function of density ratio and circulation were computed.  For a given density ratio below a limiting value, a unique shape, translating with constant velocity, was found. In such equilibria studies, the vortex sheet strength at the patch boundary is assumed constant. Evolution equations for varying sheet strength have been typically obtained in models of irrotational flows, for example, the paper by Baker and Moore \cite{baker} that computes the evolution of a buoyant air bubble, by generalising the Birkhoff-Rott equation for vortex sheet evolution in a single density fluid, and the paper by Baker, Meiron and Orszag \cite{bmo} that computes free-surface flows by modelling the free surfaces as vortex sheets; a Fredholm integral equation governing the sheet strength is derived. The jump in the tangential velocity for a buoyant patch may be contrasted with the continuity of the velocity at the boundary of a neutrally buoyant vortex patch. However, it appears possible to consider a situation of continuous velocity, with attendant pressure discontinuity, at the interface;  this possibly is sketched in \textcolor{blue}{Appendix C.}

 More recently, Ravichandran et al \cite{ravi} presented,  apart from a DNS study, a simple model for the evolution of the centroid of a buoyant circular patch, to demonstrate the `lift-induced' collapse of a vortex dipole. The evolution equation for the patch centroid (Eq.(3) of that paper) is written in an ad hoc manner. Carpenter \& Guha \cite{carp} also present an ad hoc model for buoyant point vortices. The present analysis will clarify issues relevant to these models.

 The main goal of the present work is to derive the correct equations governing an isolated buoyant vortex patch. First, applying the momentum theorem to a standard control volume, that includes the patch and surrounding fluid, the equation governing the motion of the centroid of the patch is derived. Along with the linear momentum equation, the equation governing the evolution of the  buoyant patch boundary which will, in general, deform, is analyzed. Exploiting the inherent $\operatorname{SE}(2)$ symmetry in the flow, use is made of a centroid-fixed translating frame in deriving the equations of the boundary which allows for a decomposition of the deformation and translation velocity fields. Finally, \textcolor{blue}{baroclinicity at the patch boundary creates a vortex sheet of time-varying strength.} It is shown in detail how a single integral equation for the evolution of the sheet strength can be obtained and how the combined set of equations can then, in principle, be propagated in time. To the best of our knowledge, a set of such general coupled equations governing the deformation and translation of a buoyant vortex patch of any smooth shape has not been derived before. This paper only touches upon  numerically solving these equations, a more detailed exploration alone these lines is planned for the future. But by deriving these equations, we want to exhibit the enormous complexities that enter the mathematical model of a classical vortex patch by simply relaxing the assumption of homogeneous density. Last, but not least, the point vortex limit is examined.

Note that an alternative to using a vortex sheet is to model the irrotational flows generated by the sheet using Zakharov's formulation for the motion of free surfaces in the Hamiltonian variables of the boundary position and the velocity potential function on the boundary. Such an approach is used in the problems considered in, for example, the papers \cite{Bb1987, LeMaMoRa1986}. But evolving the system involves extensive integration of Neumann or Dirichlet problems. Moreover, the boundary is not `free', and so the Zakharov formulations would entail the additional task of dealing with the pressure on the boundary. As is well-known, advantages afforded by a vortex sheet is that it avoids the explicit use of pressure and the need for integrating Laplace's equation in the domain. 

The paper is organized as follows. In Section 2, we present the linear momentum analysis for a buoyant vortex patch and derive the equation governing the motion of the vortex centroid.  In Section 3, we present the decomposition of the velocity fields and the evolution equations for the patch boundary, in both the spatially-fixed and centroid-fixed frames. In Section 4, we present the pressure continuity condition and the vortex sheet strength equation that follows. This is followed by a brief outline of an algorithmic procedure for solving all the governing equations. Finally, a brief examination of the equations in several limits, including the point vortex limit, is presented. In Section 5, some concluding remarks are made. 

To keep the paper uncluttered, most of the derivations are relegated to the Appendices. Moreover, to avoid a large number of overhead arrows or boldface symbols throughout the paper, we have chosen to represent vectors (except for unit vectors) and scalars in the same manner without any distinguishing notation.

\section{Linear momentum analysis}

   Consider a buoyant vortex patch with uniform vorticity (function) $\omega$ in a time varying domain $D_v(t) \subset \mathbb{R}^2$, with boundary $\partial D_v(t)$ and invariant area $A$. The fluid in the domain $D(t):=\mathbb{R}^2 \backslash D_v(t)$ is assumed irrotational. Moreover, the fluid is incompressible everywhere but with different densities (per unit area)  inside and outside the patch, $\rho_i$ and $\rho_o$.  The fluid is in an uniform gravitational field pointing in the $-\hat{j}$ direction. 

\begin{figure}
\centering
\includegraphics[width=3in]{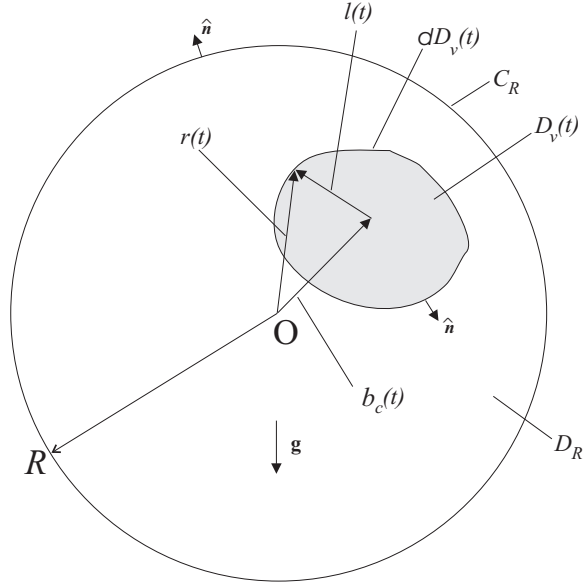}
\caption{Schematic for the buoyant vortex patch. Gravity points down.}
\end{figure}

     Now consider a fixed control volume in the shape of  a disc $D_R$ of radius $R$, centered at point $O$, with boundary $C_R$. Let $\tilde{D}(t):= D(t) \cap D_R$. Take $O$ to be the origin of a spatially-fixed frame and let $r$ denote the position vector in this frame. In addition, let $b_c(t)$ denote the position vector of the centroid of the patch in this frame, and write $r=b_c+l$. By definition of the centroid, at any time $t$, 
\[\int_{D_v(t)} l \; dA =0\]
Let $\hat{n}_v$ be the unit outward normal on the patch (i.e. pointing away from the patch) and $n$ the  unit normal on $C_R$. A schematic of the situation is shown in figure 1.

           The problem of the BVP is formulated as follows.  
   The fluid flow in $D_v$ is governed by a streamfunction obeying 
\[\nabla^2 \psi_i(r,t)=-\omega,\] 
and the corresponding velocity field $v_i=J \nabla \psi_i$, where $r$ is the position vector in a spatially-fixed frame and $J:=\left(\begin{array}{cc}0 & 1 \\-1 & 0 \end{array}\right)$. 

   The fluid flow in $D$ is governed by a velocity potential function $\phi_o$ satisfying the following Neumann problem:
\begin{align*}
\nabla^2 \phi_o &=0 \: {\rm in} \: D, \\
\nabla \phi_o \cdot \hat{n}&= v_i \cdot \hat{n} \: {\rm on} \: \partial D_v, \quad \phi_o \rightarrow {\rm constant} \: {\rm as} \: r \rightarrow \infty,
\end{align*}
and the corresponding velocity field is given by $v_0=\nabla \phi_o$. 

Applying the momentum theorem to the fluid in $D_R$ (details in Appendix A), we have

\begin{align} \label{eq:discont}
&V_c \times \rho_o \Gamma_o \hat{k}+\rho_o \frac{d}{d t}  \oint_{\partial D_v(t)}   l  \times \left( \hat{n} \times  v_o \right)\; ds  \nonumber  \\
& \hspace{1in} -  \rho_i \frac{d}{d t}  \oint_{\partial D_v(t)}   l  \times \left( \hat{n} \times  v_i \right)\; ds=(\rho_o - \rho_i) A g \hat{j}  
\end{align}
where $V_c$ is the velocity of the centroid and 
\[\Gamma_o:= \oint_{\partial D_v(t)} v_o \cdot \hat{t} ds \]
is the circulation of the outer flow. Since the outer flow is irrotational, Stokes' theorem implies that $\Gamma_o$ is the same for any closed circuit in the outer fluid that encloses the vortex patch. On the other hand, 
\[\Gamma_i:= \oint_{\partial D_v(t)} v_i \cdot \hat{t} ds=\omega A, \]
(the last equation again following from Stokes' theorem) is the circulation of the inner flow and is fixed by the patch parameters $\omega$ and $A$. By Kelvin's circulation theorem, both $\Gamma_o$ and $\Gamma_i$ are invariant in time. \textcolor{blue}{This is easily seen since $d \Gamma/dt = -\oint_\beta dp/
\rho=0$ for any closed circuit $\beta$ lying either in the outer fluid or in the inner fluid.} 

A straightforward rearrangement gives the following equation 
\begin{align}
\Gamma_o&= \omega A + \oint_{\partial D_v(t)} \gamma ds \label{eq:Gammo}
\end{align} where
\[\gamma(s,t):=(v_o-v_i) \cdot \hat{t}\] is the total slip velocity field on the patch boundary. This slip field consists of two contributions--the slip due to the vortex sheet whose strength is one of the variables of the system and another slip field generated by the instantaneous translation motion. Both these are described later. It follows that $\oint_{\partial D_v(t)} \gamma \, ds$ is also invariant in time, though $\gamma(s,t)$ is in general not. \textcolor{blue}{In other words, as in the work of \cite{bmo}, there is no global  baroclinic generation of vorticity, but the variation in $\gamma(s,t)$ may be identified with local baroclinic generation of vorticity.}

 In equation~(\ref{eq:discont}), the $dV_c/dt$ term does not appear explicitly. The role of this term is seen more clearly in the next section, where the equation is written in a different way.
 
\section{Deformations of buoyant vortex patches} 

   The natural next question to ask is how the centroid motion and the boundary deformation are linked to each other, and this is examined in this section.
 
The boundary deformation is caused by the vorticity in the patch and the essential vortex sheet which enforces the pressure continuity condition. The buoyant force on the patch causes the centroid to move and the patch to instantaneously translate which generates irrotational flows both inside and outside the patch. These flows will be modeled as Kirchhoff flows associated with the instantaneous rigid boundary shape, and are discussed more below.

  We first write the equations in a spatially-fixed frame. The boundary of the patch is  viewed as the image of smooth maps, $\partial D_v: S^1 \rightarrow \mathbb{R}^2$. Denote by $C(s,t)$ the coordinates of the image which is a smooth curve in $\mathbb{R}^2$, and by $\bar{C}(s,t)$ the same curve with respect to the translating frame defined in the next section. Assuming the patch motion at every instant to be decomposable into a rigid translation and a deformation, the fluid flow in the inner region $D_v$ is governed by the following streamfunction and velocity field
\begin{gather}
\psi_i(r,t) =\psi_{Di}(r,t) + \psi_{Ti}(r,t), \,\, 
v_i(r,t)=v_{Di}(r,t) + V_c(t), 
\label{eq:psii}
\end{gather}
where the subscripts $D, T$ and $i$ stand for deformation, translation and inner. The deformation field $\psi_{Di}$ is the sum of two components, $\psi_p$ and $\psi_{si}$, which indicate respectively the contributions of the patch and the bounding vortex sheet. Thus the various component streamfunctions are 
\begin{align*}
\psi_p(r,t)&=\frac{ \omega}{2 \pi}\int_{D_v(t)} \log \mid r - \tilde{r} \mid \; d \tilde{A}, \\
\psi_{si}(r,t)&= \frac{1}{2 \pi}  \oint_C \gamma_{s} (\tilde{s},t) \log \mid r - \tilde{r} \mid \; d \tilde{s}, \quad \tilde{r}:=r(\tilde{s}) \\
\psi_T(r,t)&=( V_c(t) \times r) \cdot \hat{k}
\end{align*}
where $\gamma_{\mathfrak{s}}(s,t)$ is the strength distribution of the vortex sheet, $s$ is the boundary curve parameter and $V_c(t)$ is the velocity of the patch centroid. $\psi_{Ti}$ is the harmonic streamfunction of the internal Kirchhoff flow generated by the instantaneous rigid body translation of the patch. It may be recalled that the velocity field of this flow is equal to $V_c(t)$ everywhere in the patch \cite{lamb}. 
     The other component velocity fields are 
\begin{align}
v_p(r,t)&=-\frac{1}{2 \pi} \omega \hat{k} \times \int_{D_v(t)}  \nabla_{r} \log \mid r - \tilde{r} \mid \; d \tilde{A}, \label{eq:vpatch}  \\
v_{si}(r,t)&=-\frac{1}{2 \pi} \oint_{C} \gamma_{\mathfrak{s}} (\tilde{s},t) \hat{k} \times  \frac{ r - \tilde{r}}{\mid r - \tilde{r} \mid^2} \; d \tilde{s}, \quad r(s) \in D_v, \label{eq:vsheet}
\end{align}

 The fluid flow in $D$ is governed by the following streamfunction and velocity field
\begin{align}
\psi_o(r,t)&=\psi_{Do}(r,t)+\psi_{To}(r,t),  \quad v_o(r,t)=v_{Do}(r,t)+v_{To}(r,t), \label{eq:psio}
\end{align}
where again the deformation field $\psi_{Do}$ is the sum of two components $\psi_p$ and $\psi_{so}$. Note that the velocity induced by a vortex patch is continuous across the boundary so that $v_p$ for points in $D$ is given by the same expression as~(\ref{eq:vpatch}). The vortex sheet velocity field expression is also the same as ~(\ref{eq:vsheet}) but recall \cite{Sa1992} that the expression leads to a jump in the tangential velocity across the boundary. The Kirchhoff flows also induce a discontinuous tangential velocity at the boundary. $\psi_{To}$ is the harmonic conjugate of $\phi_{To}$ which satisfies the following Neumann problem, 
\begin{align*}
\nabla^2 \phi_{To} &=0 \: {\rm in} \: D, \\
\nabla \phi_{To} \cdot \hat{n} (\equiv v_{To} \cdot \hat{n})&= V_c\cdot \hat{n} \: {\rm on} \: \partial D_v, \quad \phi_{To} \rightarrow {\rm constant} \: {\rm as} \: r \rightarrow \infty
\end{align*}
The above problem, when written in a body-fixed frame, is recognized as the external Kirchhoff flow generated by the instantaneous rigid body translation. Recall that this is the external irrotational incompressible flow induced by a moving rigid body. Kirchhoff \cite{Kirchhoff1869,lamb} showed that in general the velocity potential function of such a flow, referred to a body-fixed frame,\footnote{Though the body-fixed frame used by Kirchhoff is a classical concept, the problem is sometimes confused with the flow described in a non-inertial frame. In the former, each vector is expressed in terms of the instantaneous body-fixed frame and, except for the position vector, differs from the vector expressed in the spatially-fixed frame by the action of the instantaneous rotation matrix. For a purely translating frame, as in our problem, the vectors in the two frames are identical; see, for example, Goldstein \cite{Goldstein1950} for a discussion of such frames.} can be expressed in terms of the body's instantaneous velocities and unit potential functions that depend solely on the body's rigid shape. In the current problem, since the body's shape is not rigid the unit potential functions are time-dependent and, at each instant, are the unit potentials corresponding to the instantaneous shape.

    Let $\tilde{\phi}_{To}(l,t):=\phi_{To}(r,t)$, then
\begin{align*}
\tilde{\phi}_{To}(l,t)&=V_{cx}(t) a(l,\bar{C})+V_{cy}(t)b(l,\bar{C}), \\
\Rightarrow v_{To}(l,t):=\nabla \tilde{\phi}_{To}(l,t)&=E(l,\bar{C}) \cdot V_c(t)
\end{align*} where $E$ is a $2 \times 2$ matrix of the first order spatial derivatives of the unit potentials $a$ and $b$. The unit potentials  typically have analytic expressions only for simple shapes, or shapes obtained from conformal maps. For arbitrary shapes, a Laplace equation solver may have to be appended that numerically computes these functions at each instant in a neighborhood of $\bar{C}$.    
Recall that the sheet strength distribution is given by 
\[\gamma_s (s,t)=(v_{so}-v_{si}) \cdot \hat{t} \]
with $v_{so}, v_{si}$ evaluated for boundary points and given by standard vortex sheet relations \cite{Sa1992}
\begin{gather}
v_{so} = \frac{\gamma_s}{2} \hat{t} + CPV, \, v_{si} = - \frac{\gamma_s}{2} \hat{t} + CPV  \label{eq:vso}
\end{gather}
where $CPV$ denotes the Cauchy Principal Value of the contour integral in ~(\ref{eq:vsheet}) when evaluated for points on the boundary.
Keeping in mind that for the vortex sheet $v_{so}(r,t) \cdot \hat{n}=v_{si}(r,t) \cdot \hat{n}$ on $\partial D_v$, one obtains the required continuity of normal velocity at the boundary, $v_o(r,t) \cdot \hat{n}=v_i(r,t) \cdot \hat{n}$ on $\partial D_v$. 

Moreover, $\gamma$ and $\gamma_s$ are related by
\begin{align*}
 \gamma&= \gamma_{\mathfrak{s}} + [(E-I) \cdot V_c] \cdot  \hat{t}
\end{align*}
from which it follows that 
\[\oint_{\bar{C}}\gamma\; ds=\oint_{\bar{C}}\gamma_{\mathfrak{s}}\; ds,\]since the Kirchhoff flows have zero circulations associated with them. Referring to equation~(\ref{eq:Gammo}), this implies that $\Gamma_o$ is the sum of the circulations due to the vortex patch and the vortex sheet.

 The evolution of the patch boundary is given, as usual, by 
\begin{align}
\frac{\partial C}{\partial t}&=(v_i \cdot \hat{n}) \hat{n}=\left(J \nabla \psi_i \cdot \hat{n} \right) \hat{n}, \label{eq:dcdt}
\end{align}
where $J:=\left(\begin{array}{cc}0 & 1 \\-1 & 0 \end{array}\right)$. Note that only the normal component of the velocity field  contributes to the evolution of $C(s,t)$.
\subsection{Centroid-fixed translating frame} With a view to obtaining an equation for the centroid velocity $V_c,$ we now write ~(\ref{eq:discont}) and ~(\ref{eq:psii}) in a centroid-fixed frame translating parallel to the stationary frame with origin at the patch centroid. With respect to this frame, ~(\ref{eq:psii}) can be written as 
\begin{align}
\psi_i(r,t)&=\tilde{\psi}_i(l,t) \nonumber \\
&=\underbrace{\frac{ \omega}{2 \pi}\int_{D_v(t)} \log \mid l - \tilde{l} \mid \; d \tilde{A}+\frac{1}{2 \pi}  \oint_{\bar{C}} \gamma_{\mathfrak{s}} (\tilde{s},t) \log \mid l - \tilde{l} \mid \; d \tilde{s}+( V_c \times l) \cdot \hat{k}}_{\psi_i(l,t)} \nonumber \\
& \hspace{2in} +(V_c \times b) \cdot \hat{k}, \label{eq:psi_centroid}
\end{align}
$l$ being the position vector in this frame. It is important to note that for points on the boundary $l$ is exactly the same as $\bar{C}(s,t)$ the boundary curve coordinates in this frame. The last term on the right is a purely time-dependent function. The curve evolution equation in the translating frame is seen to be (details in Appendix B)
 \begin{align}
\frac{\partial \bar{C}}{\partial t}&=\bigg{\{}\left(-\frac{ \omega}{2 \pi}  \oint_{\bar{C}}   \log \mid l - \tilde{l} \mid  \tilde{\hat{t}}  \; d \tilde{s} - \frac{1}{2 \pi} \oint_{C} \gamma_{\mathfrak{s}} (\tilde{s},t) \hat{k} \times  \frac{ l - \tilde{l}}{\mid l - \tilde{l} \mid^2} \; d \tilde{s} \right) \cdot \hat{n} \bigg{\}} \hat{n} \label{eq:cevol}
\end{align}
Due to the $\operatorname{SE}(2)$ symmetries in the streamfunctions of the patch and the sheet the deformation equation in the centroid-fixed frame becomes independent of the drift of the centroid. To predict the latter one needs the linear momentum equation~(1). The $\operatorname{SE}(2)$ symmetries mean that the velocity fields due to these streamfunctions are the same relative to the patch regardless of its absolute location and orientation in the plane. 

Noting that equation~(1) is equally valid in the centroid-fixed frame, rewrite it as: 
\begin{align*} 
&V_c \times \rho_o \Gamma_o \hat{k}+\rho_o \frac{d}{d t}  \oint_{\bar{C}}   l  \times \left( \hat{n} \times ( v_o - v_i ) \right)\; ds  \nonumber  \\
& \hspace{1in} + (\rho_o-  \rho_i) \frac{d}{d t}  \oint_{\bar{C}}   l  \times \left( \hat{n} \times  v_i \right)\; ds=(\rho_o - \rho_i) A g \hat{j}  
\end{align*}
Now apply the vector identity~(A1), and another,  to write 
\begin{align*}
\oint_{\bar{C}}   l  \times \left( \hat{n} \times  v_i \right)\; ds&=\int_{D_v} l \times \omega \; dA-\int_{D_v} v_i \; dA, \\
&=\int_{D_v} l \times \omega \; dA- \hat{k} \times \int_{D_v} \nabla_l \Psi_i \; dA-  \int_{D_v} V_c \; dA, \\
&=\int_{D_v} l \times \omega \; dA- \hat{k} \times \oint_{\bar{C}}  \Psi_i  \hat{n} \; ds - V_c A, \\
&=\int_{D_v} l \times \omega \; dA-  \oint_{\bar{C}}  \Psi_i  \hat{t} \; ds - V_c A
\end{align*}
so that 
\[\frac{d}{dt} \oint_{\bar{C}}   l  \times \left( \hat{n} \times  v_i \right)\; ds =-\frac{d}{dt} \oint_{\bar{C}}\Psi_i  \hat{t} \; ds-A \frac{d V_c}{dt}\] 
the vorticity term being constant for a vortex patch. $\Psi_i$ is defined by eq.(B2). 

It follows that 
\[\oint_{\bar{C}}   l  \times \left( \hat{n} \times ( v_o - v_i ) \right)\; ds=\oint_{\bar{C}}  l  \times \gamma \hat{k} \; ds=-\hat{k} \times \oint_{\bar{C}} \gamma  l  \; ds\] and so finally the equation for $V_c$ is 
\begin{align}
V_c \times \rho_o \Gamma_o \hat{k}+(\rho_i-  \rho_o) A \frac{dV_c}{dt}&=(\rho_o-  \rho_i) \frac{d}{d t}  \oint_{\bar{C}} \Psi_i  \hat{t} \; ds + \hat{k} \times \rho_o \frac{d}{dt} \oint_{\bar{C}} \gamma  l  \; ds  \nonumber \\ 
& \hspace{2in} + (\rho_o - \rho_i) A g \hat{j} \label{eq:linmom}
\end{align}
Equations ~(\ref{eq:cevol}) and ~(\ref{eq:linmom})  are the evolution equations of the system in the variables $(\bar{C}(s), V_c)$. 

\section{Pressure continuity at the boundary}
However, to complete the system, we need an evolution equation for the sheet strength $\gamma_\mathfrak{s}(s)$. This can be derived by noting that one further condition has to be satisfied on the interface - the so-called dynamic condition, whose standard expression is the continuity of pressure across the interface. 

   With pressure continuity, the velocity at the interface must obey the following slip rule (derivation in Appendix C) 
\begin{align}
& \rho_i \left(\frac{Dv_i}{Dt} \cdot \hat{t} \right)(x(s,t),y(s,t))- \rho_o \left(\frac{Dv_o}{Dt} \cdot \hat{t} \right) (x(s,t),y(s,t)) \nonumber \\
& \hspace{2in}=-\left(\rho_i-\rho_o \right) g \hat{j} \cdot \hat{t}(x(s,t),y(s,t)) \label{eq:pressure}
\end{align}
Note that an integrated form of this equation, leading to Bernoulli equations for the inner and outer regions, has been generally used in the literature to satisfy the dynamic condition. However, we note that pressure continuity is only one of at least three possibilities. The other two are described in Appendix C as well. 

  Rewriting (\ref{eq:pressure}) as 
\begin{align}
& \left(\rho_i- \rho_o \right) \left(\frac{Dv_i}{Dt} \cdot \hat{t} \right)- \rho_o \left(\frac{Dv_o}{Dt}- \frac{Dv_i}{Dt}\right) \cdot \hat{t} \nonumber \\
& \hspace{2in}=-\left(\rho_i-\rho_o \right) g \hat{j},  \label{eq:pressure2}
\end{align}
 and plugging the forms of $v_i$ and $v_o$ from Section 3 in~(\ref{eq:pressure2}), the equation for $\gamma_s$ can be shown to be 
\begin{gather} 
\begin{aligned}\label{eq:sheetevol}
&(\rho_i - \rho_o) \left( \frac{dV_c}{dt}+ \frac{\partial}{\partial t} CPV +\frac{\partial v_p}{\partial t}+( V_c + v_p+v_{si}) \cdot \nabla v_{p}+( V_c + v_p) \cdot \nabla v_{si}  \right) \cdot \hat{t}  \\
& \hspace{2in} - (\rho_i - \rho_o) \left( \frac{1}{2} \frac{\partial \gamma_s}{\partial t} - \frac{\partial}{\partial s} \left(\frac{v_{si}^2}{2} \right) \right)  \\
&- \rho_o \bigg{(}(E-I) \cdot \frac{dV_c}{dt}+[(E-I) \cdot V_c+v_{so}-v_{si}] \cdot \nabla v_{p} +( E
 \cdot V_c + v_p) \cdot \nabla v_{so}  \\
& \hspace{0.5in} +\mathbf{D}_t E \cdot V_c+ \left(E \cdot V_c+v_p+v_{so} \right) \cdot \nabla (E \cdot V_c)-  (V_c+v_p) \cdot \nabla v_{si} \bigg{)} \cdot \hat{t}  \\ & - \rho_o \left( \frac{\partial \gamma_s}{\partial t} + \frac{\partial}{\partial s} \left(\frac{v_{so}^2}{2}\right) -v_{si} \frac{\partial}{\partial s} \left(\frac{v_{si}^2}{2}\right) \right) = (\rho_o - \rho_i) g \hat{j}\cdot \hat{t}.
\end{aligned}
\end{gather}
The notation $\mathbf{D}_tE$ is explained in Appendix D, after the equation for $d/dt(\oint_{\bar{C}} \gamma l \; ds)$. This term can be estimated from the updated boundary shape obtained from (\ref{eq:cevol}).
Referring to equations~(\ref{eq:linmom}) and~(\ref{eq:sheetevol}), they contain the first order time derivatives of six terms: $V_c, \gamma_{\mathfrak{s}}, v_p, \oint_{\bar{C}} \Psi_i  \hat{t} \; ds, \oint_{\bar{C}} \gamma l \; ds$ and $CPV$.  Even when combined with equation~(\ref{eq:cevol}) it would appear impossible to find an algorithm to propagate solutions in time. However, it is shown in Appendix  D that using~(\ref{eq:cevol}), the time derivatives of the last four terms consist of terms that depend on (a) known terms at the current  time $t$, (b) $dV_c/dt$ or (c) $\partial \gamma_{\mathfrak{s}}/\partial t$. This allows one to transform equations~(\ref{eq:linmom}) and~(\ref{eq:sheetevol}) such that they contain only two unknowns at $t$, namely, the time derivatives of $V_c$ and $\gamma_{\mathfrak{s}}$. A further elimination leads to a single integral equation containing $\partial \gamma_{\mathfrak{s}}/ \partial t$. The procedure is outlined below.

Importing terms from Appendix D, we can rewrite~(\ref{eq:linmom}) as
\begin{align}
&\frac{d V_c}{dt} \nonumber \\
&= \bigg{[}A \left( \rho_i - \rho_o \right) I +B \bigg{]}^{-1} \nonumber \\
& \hspace{0.2in} \cdot  \bigg{[} \left( \rho_o - \rho_i \right) \left[ W \left(\frac{\partial \gamma_{\mathfrak{s}}}{\partial t},t \right) + A g \hat{j} \right] + \hat{k} \times \rho_o X \left(\frac{\partial \gamma_{\mathfrak{s}}}{\partial t},t \right) - V_c \times \rho_o \Gamma_o \hat{k} \bigg{]} \label{eq:dvcdt}
\end{align}
where $B$ is another  $2 \times 2$ matrix containing loop integrals of combinations of elements of $l$ and $(E-I)\cdot\hat{t}$. Starting from an initial  choice of $\bar{C}(s,t), V_c(s,t)$ and $\gamma_{\mathfrak{s}}(s,t)$, $\partial \bar{C}/ \partial t(s,t)$ is first obtained from~(\ref{eq:cevol}) which immediately produces an estimate of $\bar{C}(s,t+\triangle t)$. Next, substituting for $dV_c / dt$ from~(\ref{eq:dvcdt})and the time derivative terms from Appendix D in~(\ref{eq:sheetevol}), we obtain a single integral equation which can, in principle, be solved for the only unknown $\partial \gamma_{\mathfrak{s}} / \partial t (s,t).$ Using these in~(\ref{eq:dvcdt}), produces $dV_c / dt$ at time $t$. With these time derivatives known at time $t$, $V_c(t+\Delta t)$ and $\gamma_{\mathfrak{s}}(s,t+\Delta t)$ are obtained and the procedure is repeated.

We now check for various special cases. For a neutrally buoyant patch ($\rho_i = \rho_o$), pressure and velocity are continuous on the boundary and hence $\gamma = 0,$ which by~(\ref{eq:linmom}) implies $V_c = V_c(0).$ The vortex patch evolution then reduces to a solution of (\ref{eq:cevol}), as it should (for e.g. \cite{Sa1992}). Other special solutions are the steady states of \cite{kao}. Putting all time derivatives to zero in~(\ref{eq:linmom}) readily gives 
\begin{align}
V_c &= \frac{(\rho_o - \rho_i) A g}{\rho_o \Gamma_o} \hat{i} \label{eq:vel_centroid}
\end{align}
This expression is same as in \cite{kao}. The fact that $V_c$ has a simple expression, independent of the complicated calculations to find $l$ and $\gamma_{\mathfrak{s}}$, just reflects the fact that the patch motion can be decomposed into a rigid translation and a deformation.

\paragraph{Numerical simulations of some vorticity-free cases.} A basic numerical code using MATLAB was developed to simulate some simple cases, in all of which the patch has zero vorticity. This allowed us to compare our results with that of Baker and Moore's simulations for the evolution of  an initially circular patch with zero vorticity, and we chose the same parameter values $\rho_o = 1, \rho_i = 0$ and $g = 1$ \cite{baker}. It may be recalled that their motivation was to simulate the experimentally observed phenomenon of a jet forming in the rear of an almost planar buoyant gas bubble.  We also considered two cases in which the initial shape is an ellipse of small eccentricity. In all the three cases, the shapes are initially symmetric about the $y$-axis and consequently maintain this symmetry during evolution. Therefore, we solve only for the left side and obtain the right side by symmetry. As described above, in our method we have to solve the coupled system (\ref{eq:linmom}) and (\ref{eq:sheetevol}), while updating the boundary at each time step using (\ref{eq:cevol}). Since the centroid moves only in the vertical direction, $V_c$ is  viewed as a scalar, and the sheet strength is discretised as $\gamma_{\mathfrak{s}} = \sum_{n=1}^{N} f_n(t) \sin n \theta$, where $\theta \in [0, \pi]$ is reckoned positive counterclockwise from the positive $y$-axis. 

  We need a system of $(N+1)$ ODEs for the $(N+1)$ time dependent unknowns $V_c, f_i, i = 1, ...N$. The first of these is generated by plugging in the forms of $V_c$ and $\gamma_s$ into (\ref{eq:linmom}); it is seen that the horizontal components vanish.  The remaining $N$ equations of the ODE system are obtained by plugging the forms of $V_c$ and $\gamma_{\mathfrak{s}}$ into (\ref{eq:sheetevol}) which yields an expression of the form
$\sum_{n=1}^{N} C_n(dV_c/dt, d \gamma_{\mathfrak{s}} / dt) \sin n \theta = 0$, where the $C_n$ are linear in the unknowns; for simplicity, we have not displayed the $C_n$. This is possible only if $C_n = 0$ for $n = 1, ...N$ and leads to the remaining $N$ ODEs.

\begin{figure}[t]
\centering
\includegraphics[width=3in]{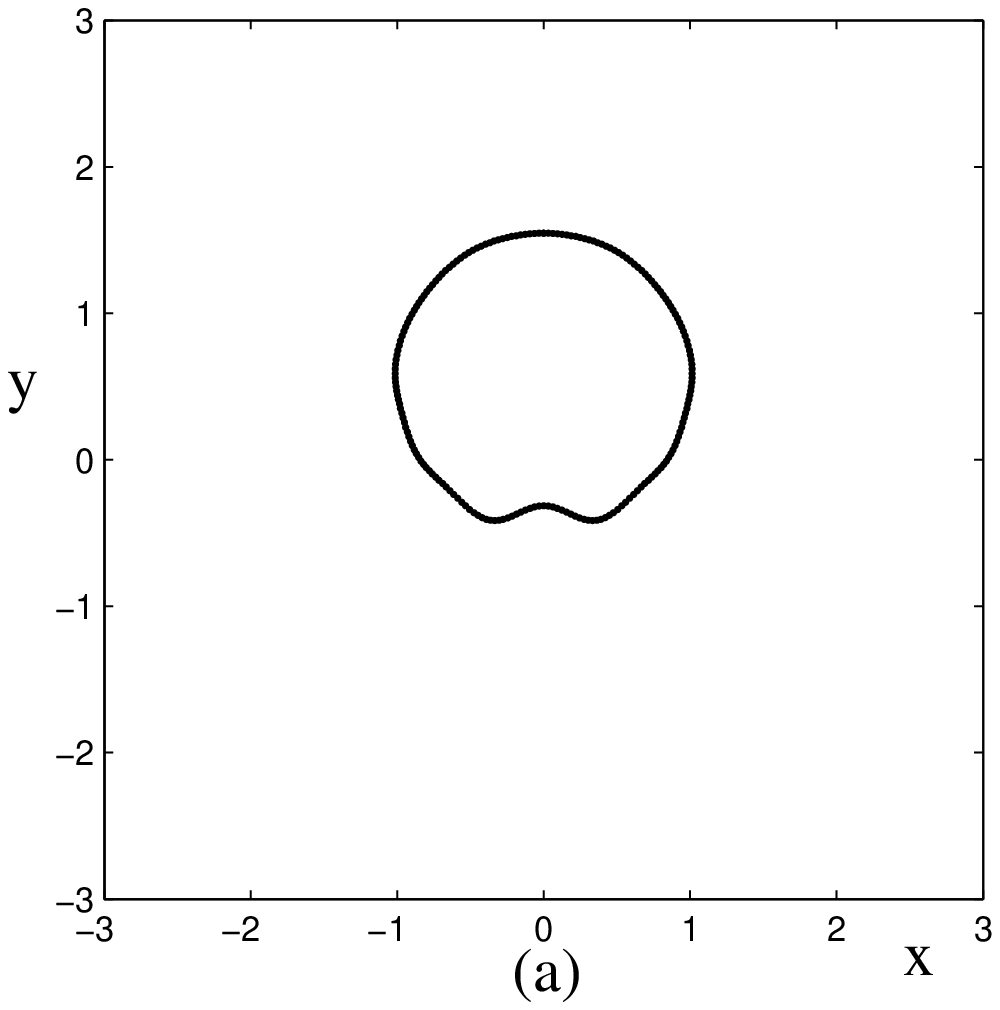}
\includegraphics[width=3in]{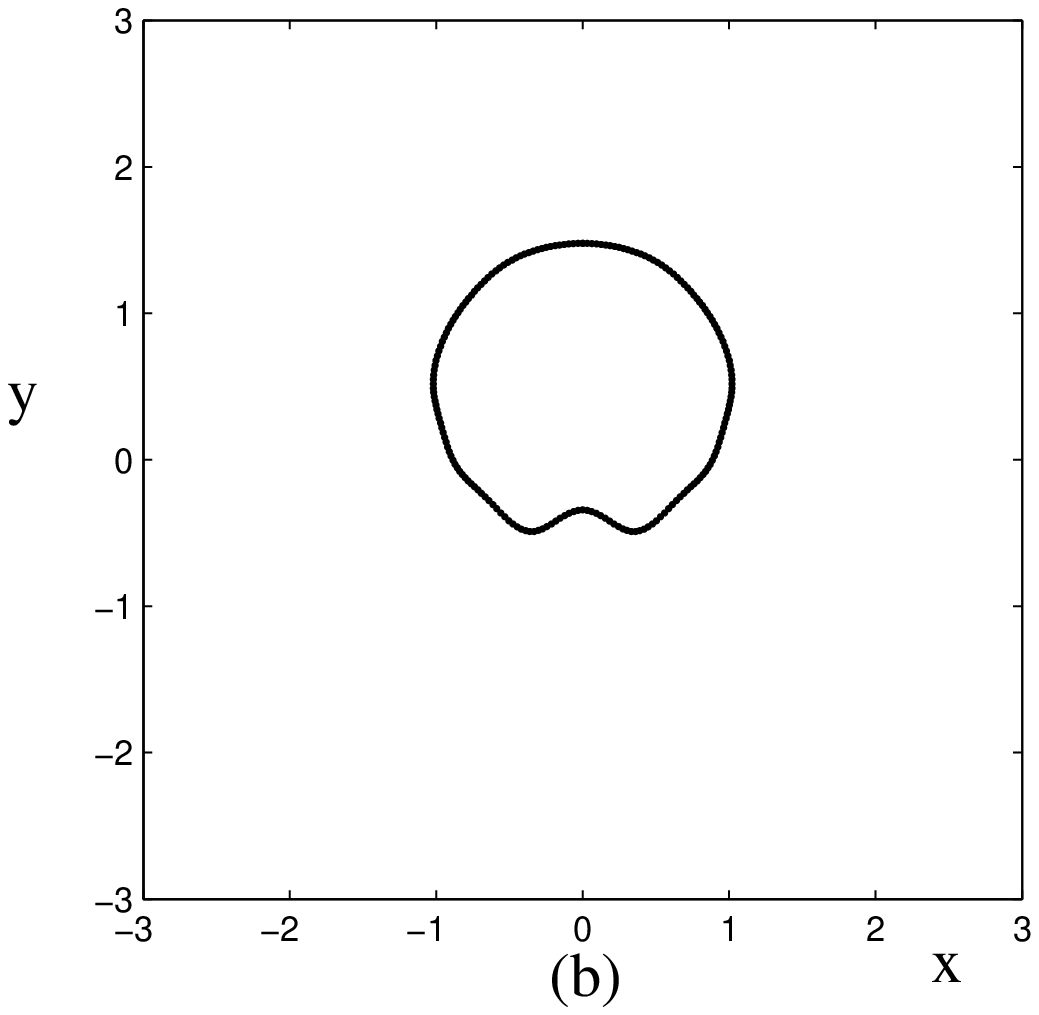}
\includegraphics[width=3in]{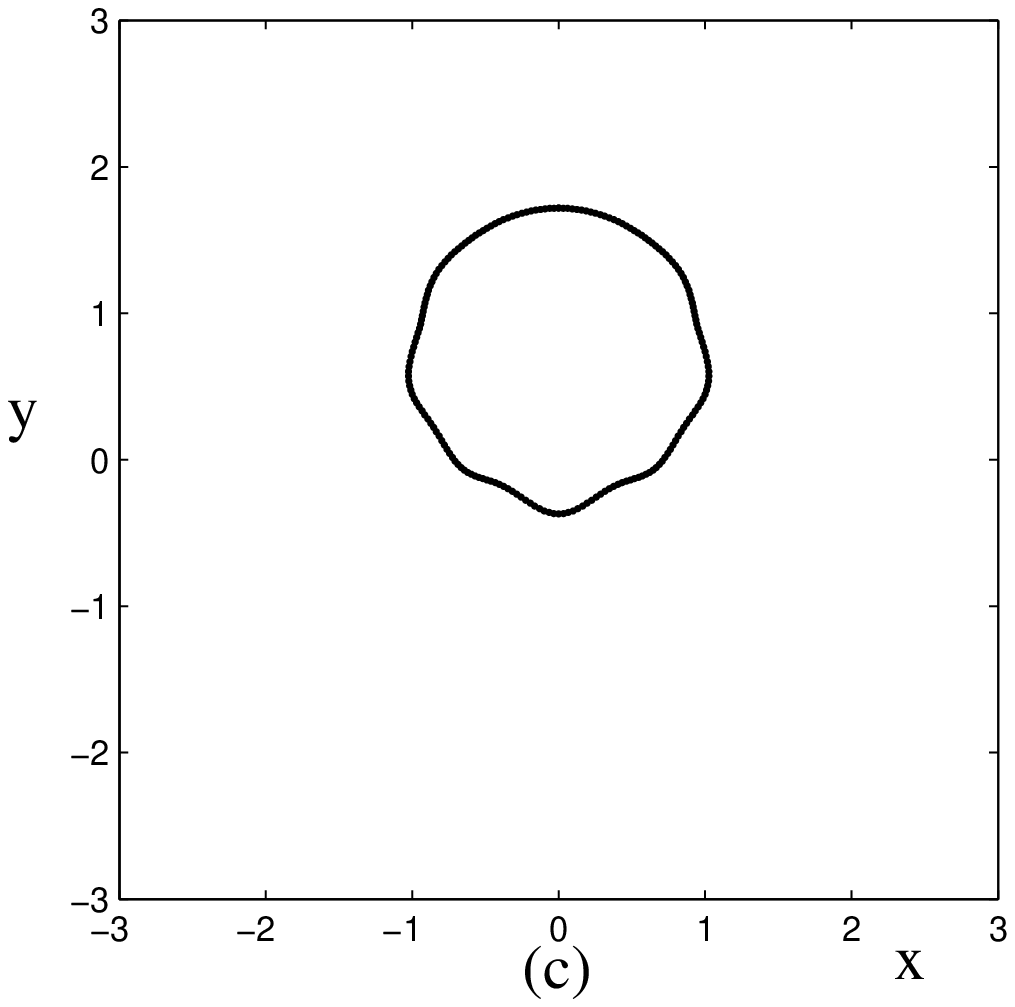}
\caption{Evolution of the patch boundary, separating fluids of densities $\rho_o = 1$ and $\rho_i = 0$ with different initial conditions. At time $t=0$ the centroid of the bubble is at $x=y=0$.  Acceleration due to gravity $g=1.$ (a) Initially circular boundary with $\gamma_s = 10^{-5} \sin \theta$, at $t = 1.08,$ (b) initially elliptic boundary with $a=1.001, b=1, \gamma_s = 0$, at $t = 1.02$ and (c) initially elliptic boundary with $a = 1, b = 1.001, \gamma_s = 0$, at $t = 1.20$. Note the incipient upward jet in the first two cases.}
\label{sim}
\end{figure}

     Denoting the vector $u = {V_c,f_1,..f_N}$, the ODE system is finally written as $A du/dt = B$, which is then integrated by a simple first order Euler time stepping. Updating the boundary position of the boundary via (\ref{eq:cevol}) is also by Euler time stepping. Computations can be continued till a time the patch boundary remains single valued in $\theta$. Results are presented  in Figure \ref{sim} for three different cases of initial conditions- (a) circular boundary with $\gamma_{\mathfrak{s}} = 10^{-5} \sin \theta$, (b) elliptic boundary with semi-major and semi-minor axes $a = 1.001, b = 1.$ with $\gamma_{\mathfrak{s}}=0$ and (c) elliptic boundary with $a = 1, b = 1.001$ with $\gamma_{\mathfrak{s}}=0$. 100 boundary points and 10 fourier modes were used in the computation. For the Baker and Moore case, Figure \ref{sim} (a), a dimple suggesting an incipient jet formation is observed. However, as soon as folds develop numerical instability sets in. The comparison with \cite{baker} was therefore mostly inconclusive.

    The procedure is similar for non-zero patch vorticity. However, there will in general be no symmetry and the full fourier series for $\gamma_s$ will have to be used. The centroid vorticity will  have two components $V_{cx}$ and $V_{cy}$. 

\subsection{Small buoyant patches and buoyant point vortices}
In the limit of vanishing patch size, the equations derived in the previous sections should lead to the equation governing an isolated buoyant point vortex and thereby to the equations governing interacting buoyant point vortices. For $\epsilon$ patches (small buoyant patches of characteristic length $\epsilon$), and especially in the limit of vanishing size, equations~(\ref{eq:cevol}) and~(\ref{eq:sheetevol}) are irrelevant. 

 Examining~(\ref{eq:linmom}) in the limit of $\epsilon \rightarrow 0$, it is reasonable to assume that the sheet strength $\gamma_{\mathfrak{s}}$, and all other slip velocity fields, go to zero whereas $\omega \rightarrow \infty $ such that $\omega A = constant=\Gamma_o$. One can write $\Psi_i \mid_{\bar{C}}=\Psi_i(0)+O(1)$, where $\Psi_i(0)= O(\log \epsilon)$ is the value at the centroid and corresponds to the value for a point vortex. It follows that $\oint_{\bar{C}} \Psi_i \hat{t} \; ds=O(\epsilon^2)$, and $\oint_{\bar{C}} \gamma l \;ds$ goes to zero even faster. The time scale in the limit should correspond to that of the centroid motion, i.e. $O(1)$, and so the time derivatives of these integrals should be of the same order as the integrals. We now distinguish two possibilities.  
\begin{itemize}
\item $\rho_i \rightarrow \infty $ such that $\rho_i A = constant$ where $A= O(\epsilon^2).$ This is the so-called massive point vortex \cite{glass}. In this limit, the integrals with $\rho_o$ as coefficient are negligible, being $O(\epsilon^2)$. However, conistent with the above estimates, the term consisting of the integral involving $\Psi_i$, with $\rho_i$ as coefficient, is $O(1)$,  which renders it non-negligible. The dynamics of the point vortex is governed by a second-order differential equation in its position coordinates, involving the gyroscopic Kutta lift and the weight of the vortex. Motion of the point vortex in this case is possible. 
\item $\rho_i = const.$ This is the classical point vortex system, valid for homogeneous fluid. The dynamics is governed by the well-known standard first order ODE system, and the vortex will remain at rest.
\end{itemize}
Glass et al \cite{glass} examined the limiting dynamics when a rigid body with fixed circulation shrinks to a point and highlighted that this limit leads to a singular perturbation problem which has to correctly analysed. They purport to provide such an analysis and demonstrate the above mentioned two distinguished limits of the massive and classical point vortices. A particularly interesting aspect of the problem is the dependence of the singularities on the symmetry of the shrinking body and their detailed treatment. A similar detailed analysis for the present model may be required to obtain the correct buoyant point vortex equations. For another perspective on point vortices with mass, see \cite{RaKoOl1994}.

\section{Concluding remarks}
The evolution of a two-dimensional buoyant vortex patch is studied  by a clear and detailed momentum analysis of its motion. Framing the motion as being due to motion of the patch centroid and a deformation results in a rational derivation of equations of motion for the centroid and the patch boundary. It is seen that the centroid motion depends on the internal flow in the patch as well as the patch shape. A similar, but simpler, equation for a vortex with a finite, but small, buoyant circular core, was given (eq. A4, \cite{Sa1972}) in an ad hoc manner; core deformation and velocity variations in the core were neglected. Ravichandran et al (2017) \cite{ravi} also write an ad hoc equation for the centroid of a buoyant Rankine vortex which is incorrect in fundamental ways, with even the apparent mass term $\rho_o d V_c / dt$ omitted; the effect of the vortex sheet and its deformation is anyway not included. This leads to the incorrect claim (eq.(5) of \cite{ravi}) that a Rankine vortex, i.e. a patch of circular shape, translates at constant speed for any density ratio; this is in direct contradiction of the results obtained in \cite{kao} and in this paper. In fact, the equation purportedly describing the motion of the vortex centroid, eq.(3) in \cite{ravi}, is really a description of that of a buoyant rigid cylinder with non-zero circulation. 

The present analysis, from first principles, demonstrates conclusively the effects of the added mass and the  deformation of the vortex patch --due to the patch vorticity and the essential slip field resulting from enforcing pressure continuity-- on the motion of the centroid.   The analysis is facilitated by splitting the velocity field into ones due to the vortex patch, a bounding vortex sheet and the internal and external Kirchhoff flows generated by the translation of the instantaneous patch shape. This decomposition lays bare the physical causal chain, which can be summarised as a) the velocity fields due to the patch and bounding vortex sheet causing a change in shape of the patch given by~(\ref{eq:cevol}), which then, by ~(\ref{eq:linmom}) and~(\ref{eq:sheetevol}), (b) leads to a change of the sheet strength distribution $\gamma_s$ and the centroid velocity $V_c$  which then (c) changes the patch shape (again through~(\ref{eq:cevol})). Even though the evolution equations for $\gamma_s$ and $V_c$ appear to be coupled in a complicated manner, it turns out to be possible to express $d V_c / dt$ in terms of $\partial \gamma_s / \partial t$ and $d l / dt.$  The decoupling allows the computation of the patch shape, sheet strength and centroid velocity in that order. The analysis also reveals that the point vortex limit should be carefully carried out. Another interesting direction would be to construct a Hamiltonian model of interacting buoyant patches, along the lines of \cite{MeZaSt1986}.

\section*{Appendix A }

    The momentum theorem applied to the fluid in $D_R$, at a given time $t$, gives 
\begin{align*}
\rho_i \frac{d}{d t} \int_{D_v(t)} v \; dA + \rho_o \frac{d}{d t} \int_{\tilde{D}(t)} v \; dA&=-\oint_{C_R} p \hat{n} \; ds-\left(\rho_i A g + \rho_o(\pi R^2 - A) g\right)\hat{j},
\end{align*}
Now, use the vector identity valid for any vector field $a$ in a domain $\mathfrak{D} \subset \mathbb{R}^2$,
\begin{align}
\int_{\mathfrak{D}} a \; dA &= \int_{\mathfrak{D}}  r \times (\nabla \times a) \; dA - \oint_{\partial \mathfrak{D}}  r \times \left( \hat{n} \times  a \right)\; ds \tag{A1}
\end{align}
to write the momentum theorem as 
\begin{align*}
& \rho_i \frac{d}{d t} \left(\int_{D_v(t)}  r \times \omega \hat{k}   \; dA - \oint_{\partial D_v(t)}  r \times \left( \hat{n} \times  v_i \right)\; ds \right) \\
& + \rho_o \frac{d}{d t} \left(\int_{D(t)}  r \times \omega \hat{k}   \; dA + \oint_{\partial D_v(t)}  r \times \left( \hat{n} \times  v_o \right)\; ds - \oint_{C_R}  r \times \left( \hat{n} \times  v_o \right)\; ds \right) \\
& \hspace{2in} =-\oint_{C_R} p \hat{n} \; ds-\left(\rho_i A g + \rho_o(\pi R^2 - A) g\right)\hat{j}, \\ \\
\Rightarrow & \rho_i \frac{d}{d t} \int_{D_v(t)}  r \times \omega \hat{k}   \; dA + \rho_o  \frac{d}{d t}  \oint_{\partial D_v(t)}  r \times \left( \hat{n} \times  v_o \right)\; ds - \rho_i \frac{d}{d t}  \oint_{\partial D_v(t)}  r \times \left( \hat{n} \times  v_i \right)\; ds\\
& \hspace{1in} =   \rho_o \frac{d}{d t} \oint_{C_R}  r \times \left( \hat{n} \times  v_o \right)\; ds  -\oint_{C_R} p \hat{n} \; ds-\left(\rho_i A g + \rho_o(\pi R^2 - A) g\right)\hat{j}, \\ \\
\Rightarrow & \rho_i \frac{d}{d t} \int_{D_v(t)}   (b_c(t) + l)  \times \omega \hat{k}   \; dA + \rho_o  \frac{d}{d t}  \oint_{\partial D_v(t)}   (b_c(t) + l)  \times \left( \hat{n} \times  v_o \right)\; ds \\
& \hspace{0.7in} - \rho_i \frac{d}{d t}  \oint_{\partial D_v(t)}   (b_c(t) + l)  \times \left( \hat{n} \times  v_i \right)\; ds \\
& \hspace{1.2in} =   \rho_o \frac{d}{d t} \oint_{C_R}  r \times \left( \hat{n} \times  v_o \right)\; ds  -\oint_{C_R} p \hat{n} \; ds-\left(\rho_i A g + \rho_o(\pi R^2 - A) g\right)\hat{j}
\end{align*}
Noting that, in general,  for a velocity field on any closed curve $C$
\[\oint_{\partial C} \hat{n} \times  v \; ds = \Gamma \hat{k}, \]
where $\Gamma$ is the circulation,
obtain
\begin{align*}
& V_c \times \rho_o \Gamma_o \hat{k} +\rho_o  \frac{d}{d t}  \oint_{\partial D_v(t)}   l  \times \left( \hat{n}_v \times  v_o \right)\; ds - \rho_i \frac{d}{d t}  \oint_{\partial D_v(t)}   l  \times \left( \hat{n}_v \times  v_i \right)\; ds\\
& \hspace{1in} =\rho_o \frac{d}{d t}\oint_{C_R} r \times \left( \hat{n} \times  v_o \right) \; ds  -\oint_{C_R} p \hat{n}  \; ds-\left((\rho_i - \rho_o) A + \rho_o \pi R^2  \right)g\hat{j}, 
\end{align*}
where 
\[V_c=\frac{db_c}{dt},\]
and $\Gamma_o$ is the constant circulation associated with the outside fluid (see Sections 2 and 3). 
Now use Bernoulli's on $C_R$, 
\[p=-\rho_o \left( g y + \frac{\partial \phi}{\partial t} + \frac{1}{2} \nabla \phi \cdot \nabla \phi\right)+p_\infty,\]
where $y$ is measured with respect to some datum and $p_\infty$ is a reference pressure. Evaluating just the gravity term, 
\begin{align*}
\rho_o g \oint_{C_R} y \hat{n}  \; ds&= \rho_o g \int_D \nabla y \; dA=\rho_o g \pi R^2 \hat{k}
\end{align*}
The momentum equation now reads
\begin{align*}
& V_c \times \rho_o \Gamma_o \hat{k} +\rho_o  \frac{d}{d t}  \oint_{\partial D_v(t)}   l  \times \left( \hat{n}_v \times  v_o \right)\; ds - \rho_i \frac{d}{d t}  \oint_{\partial D_v(t)}   l  \times \left( \hat{n}_v \times  v_i \right)\; ds\\
& \hspace{1in} =(\rho_o - \rho_i) A g \hat{j} + \rho_o \frac{d}{d t}\oint_{C_R} r \times \left( \hat{n} \times  v_o \right) \; ds  + \rho_o \oint_{C_R} \left( \frac{\partial \phi}{\partial t} + \frac{1}{2} \nabla \phi \cdot \nabla \phi \right) \hat{n}  \; ds
\end{align*}

\paragraph{Far-field terms.} The next step is to show that all the $C_R$ contour integral terms go to zero as $R \rightarrow \infty$. 
If $\Gamma_o \neq 0$, $\phi_o$ is multiple-valued and  the decay rates are given by 
\[\phi_o(r,\theta) \sim \Gamma_o \theta, \: \theta=\tan^{-1} \left(\frac{y-y_c}{x-x_c}\right)  \Rightarrow  v_o(r, \theta)=\nabla \phi_o(r,\theta) \sim \frac{\Gamma_o}{r}, \quad r \rightarrow \infty.\] 
 If $\Gamma_o=0$, then the decay rates are one order faster. The integral term containing $\nabla \phi \cdot \nabla \phi$ goes to zero from the far-field behavior  (in either case). Next, use another vector identity 
\[\oint_{C_R} r \times \left( \hat{n} \times  v \right) \; ds = - \oint_{C_R} \phi \hat{n} \; ds,\]
from which it is easily seen that the remaining integral terms on $C_R$ cancel leading to the final linear momentum balance equation
\begin{align*}
&V_c \times \rho_o \Gamma_o \hat{k}+\rho_o \frac{d}{d t}  \oint_{\partial D_v(t)}   l  \times \left( \hat{n}_v \times  v_o \right)\; ds  \nonumber  \\
& \hspace{1in} -  \rho_i \frac{d}{d t}  \oint_{\partial D_v(t)}   l  \times \left( \hat{n}_v \times  v_i \right)\; ds=(\rho_o - \rho_i) A g \hat{j}  
\end{align*}

\section*{Appendix B}

Noting that the $\nabla$ operator, in the respective coordinates, has the same form in the spatially-fixed and the centroid-fixed frames, the following relations hold
\[v_i=J \nabla \psi_i(r,t)=J \nabla \tilde{\psi}_i(l,t)=J \nabla \psi_i(l,t)\]
All the velocity decompositions and the velocity expressions in Section 3 continue to hold with the position vectors $r, \tilde{r}$ replaced by $l,\tilde{l}$. 
Moreover, the curve evolution in the two frames obeys the relation 
\begin{align*}
\frac{\partial C}{\partial t}&=\frac{\partial \bar{C}}{\partial t}+V_c \cdot \hat{n}_{\mid_{\bar{C}}} 
\end{align*}
 so that the curve evolution equation in the translating frame becomes 
 \begin{align*}
\frac{\partial \bar{C}}{\partial t}&=\left(J \nabla \psi_i - V_c \right) \cdot \hat{n}_{\mid_{\bar{C}}}. 
\end{align*}
This equation can also be written as  
\begin{align}
\frac{\partial \bar{C}}{\partial t}&=J \nabla \Psi_i \cdot \hat{n}_{\mid_{\bar{C}}}=\bar{v}_i \cdot \hat{n}_{\mid_{\bar{C}}}, \tag{B1}
\end{align}
where 
\begin{align}
\Psi_i(l,t)&=\frac{ \omega}{2 \pi}\int_{D_v(t)} \log \mid l - \tilde{l} \mid \; d \tilde{A} + \frac{1}{2 \pi}  \oint_{\bar{C}} \gamma_{\mathfrak{s}} (\tilde{s},t) \log \mid l - \tilde{l} \mid \; d \tilde{s} \tag{B2}
\end{align}

  As for neutrally buoyant patches, using the standard theorems of integral calculus, the velocity field due to the patch can be written as a contour integral: 
\begin{align*}
J \nabla \left( \frac{ \omega}{2 \pi}\int_{D_v(t)} \log \mid l - \tilde{l} \mid \; d \tilde{A}\right)&=-\frac{ \omega \hat{k}}{2 \pi}  \times \int_{D_v(t)}  \nabla_{l}  \log \mid l - \tilde{l} \mid   \; d \tilde{A} \\
&=-\frac{ \omega \hat{k}}{2 \pi}  \times \oint_{\bar{C}}   \log \mid l - \tilde{l} \mid  \tilde{\hat{n}}  \; d \tilde{s},\\
 &=-\frac{ \omega}{2 \pi}  \oint_{\bar{C}}   \log \mid l - \tilde{l} \mid  \tilde{\hat{t}}  \; d \tilde{s}
\end{align*}

This allows (B1) to be written in the contour integral form ~(\ref{eq:cevol}).

\section*{Appendix C}
Let $v$ denote the Eulerian velocity fields, $V$ denote the material/Lagrangian velocities and $p$ the Eulerian pressure fields, with corresponding subscripts for inner and outer fluid. Denote by $x(s,t),y(s,t)$ the $\mathbb{R}^2$-coordinates of material points on the interface, where $s$ is the interface curve  parameter. From the equivalence of the Eulerian and Lagrangian descriptions,
\[v(x(s,t),y(s,t))=V(s,t).\]Let $g$ denote the gravity vector.  Apply Euler's equation at the interface for the inner and the  outer fluid:
\begin{align*}
\rho_i \frac{Dv_i}{Dt}(x(s,t),y(s,t))&= -\nabla p_i(x(s,t),y(s,t)) - \rho_i g \hat{j}, \\
\rho_o \frac{Dv_o}{Dt}(x(s,t),y(s,t))&=-\nabla p_o (x(s,t),y(s,t)) - \rho_o g \hat{j}, \\
\end{align*}
where all spatial derivatives are one-sided derivatives. If $\hat{t}(s,t)$ denotes the unit tangent field then 
\begin{align*}
&\rho_i \left(\frac{Dv_i}{Dt} \cdot \hat{t} \right)(x(s,t),y(s,t)) \\
& \hspace{1in} =-\left(\nabla p_i \cdot \hat{t} \right) (x(s,t),y(s,t)) - \rho_i g \hat{j}\cdot \hat{t}(x(s,t),y(s,t)), \\
& \rho_o \left(\frac{Dv_o}{Dt} \cdot \hat{t} \right)(x(s,t),y(s,t))\\
&\hspace{1in} =-\left(\nabla p_o \cdot \hat{t} \right) (x(s,t),y(s,t)) - \rho_o g  \hat{j}\cdot \hat{t}(x(s,t),y(s,t)), \\
\end{align*}

\paragraph{Continuity of velocity.} If the velocity at the interface is continuous, then at each $s$
\[v_i(x(s,t),y(s,t))=v_o(x(s,t),y(s,t)), \quad \forall t\]
and, in the Lagrangian description, 
\[V_i(s,t)=V_o(s,t), \quad \forall t\]
From this equation, it follows that at each $s$, 
\[\frac{dV_i}{dt}(s,t)=\frac{dV_o}{dt}(s,t), \quad \forall t\]
Reverting to the Eulerian framework, the equivalence gives
\[\frac{Dv_i}{Dt}(x(s,t),y(s,t))=\frac{Dv_o}{Dt}(x(s,t),y(s,t)), \quad \forall t\]
\paragraph{Continuity of pressure.} If the pressure at the interface is continuous, then at each $s$
\begin{align*}
p_i(x(s,t),y(s,t))&=p_o(x(s,t),y(s,t)), \quad \forall t, \\
\Rightarrow P_i(s,t)&=P_o(s,t), \quad \forall t, \\
\Rightarrow \frac{\partial P_i}{\partial s}&=\frac{\partial P_o}{\partial s}, \quad \forall t, \\
\Rightarrow \left(\nabla p_i \cdot \hat{t} \right)(x(s,t),y(s,t))&=\left(\nabla p_o \cdot \hat{t} \right)(x(s,t),y(s,t)), \quad \forall t,
\end{align*}

\begin{prop}
For a buoyant patch, pressure and velocity continuity at the interface cannot be simultaneously satisfied. 
\end{prop}
{\bf{Proof}} \\
{\rm Proof by contradiction. If they can be simultaneously satisfied, then at the interface we have 
\begin{align*}
\left(\rho_i-\rho_o\right) \left(\frac{Dv_i}{Dt} \cdot \hat{t} \right)(x(s,t),y(s,t))&=-\left(\rho_i-\rho_o\right)g \hat{j} \cdot \hat{t}(x(s,t),y(s,t))
\end{align*}There are only two ways in which this equation can be satisfied: (i) for a neutrally buoyant patch or (ii) if acceleration of interface elements is equal to $-g \hat{j}$ plus a normal component. The latter is impossible since it implies that the patch is free falling while at the same time interface elements are traveling at constant speeds around the interface (with a normal centripetal acceleration) $\blacksquare$}

  The conclusion is therefore one can {\it choose} to make either the pressure or the velocity continuous at the interface for a buoyant patch. Choosing one  however implies a rule for the other, to satisfy Euler's equation. With pressure continuity, the velocity at the interface must obey the following slip rule: 
\begin{align*}
& \rho_i \left(\frac{Dv_i}{Dt} \cdot \hat{t} \right)(x(s,t),y(s,t))- \rho_o \left(\frac{Dv_o}{Dt} \cdot \hat{t} \right) (x(s,t),y(s,t)) \\
& \hspace{2in}=\left(\rho_o-\rho_i \right) g \hat{j}\cdot \hat{t}(x(s,t),y(s,t))
\end{align*}
This slip rule will affect the motion of the centroid and the motion of the interface as well.

Continuity of pressure is the commonly used dynamic boundary condition at surface-tension-free interfaces. Alternatively, one could argue that it is the {\it force due to pressure} that must be continuous. It is a basic fact that the force due to pressure  is equal to $\nabla p (x,y) \delta \mathcal{V}$ for a material element of volume $\delta \mathcal{V}$ centered at point $(x,y)$. 

\paragraph{Continuity of force per unit volume (due to pressure)} If the  force per unit volume (due to pressure) at the interface is continuous, then at each $s$,
\begin{align*}
 \nabla p_i (x(s,t),y(s,t))&=\nabla p_o  (x(s,t),y(s,t)), \quad \forall t,
\end{align*}
It is easy to see that we then have a similar Proposition.
 
\begin{prop} 
For a buoyant patch, force per unit volume (due to pressure) and velocity continuity at the interface cannot be simultaneously satisfied. 
\end{prop}
With force( due to pressure) continuity, the velocity at the interface, both tangential and normal components, must obey the following stronger rule: 
\begin{align*}
\rho_i \frac{Dv_i}{Dt} (x(s,t),y(s,t))- \rho_o \frac{Dv_o}{Dt}  (x(s,t),y(s,t))&=\left(\rho_i-\rho_o \right) g (x(s,t),y(s,t))
\end{align*}
Thus the equality of the gradient of pressure reveals another interesting possibility, due to taking one-sided derivatives. One can have equality of $\nabla p$ at the interface, yet the pressure itself could be discontinuous. 

\section*{Appendix D}
With $s$ denoting the arc-length parameter, $s$ and $t$ are independent variables, and we make the following identifications. 
For points on the boundary,
\[l \equiv l(s,t) \equiv \bar{C}(s,t), \quad \tilde{l} \equiv l(\tilde{s},t), \quad \hat{t} \equiv \hat{t}(s,t), \quad \tilde{\hat{t}} \equiv \hat{t}(\tilde{s},t). \]  Moreover, \[\hat{t}=\frac{\partial \bar{C}}{\partial s} \] The derivatives of the integral terms are now evaluated as follows:
\begin{align*}
&\frac{d}{d t}  \oint_{\bar{C}} \Psi_i  \hat{t} \; ds= \oint_{\bar{C}} \left(\frac{\partial  \Psi_i }{ \partial t}\hat{t}+\Psi_i \frac{\partial  \hat{t} }{ \partial t} \right) \; ds, \\
& \hspace{-0.5in}=\oint_{\bar{C}}\left( \frac{ \omega \hat{t}}{2 \pi}\frac{\partial l}{\partial t}\cdot  \int_{D_v(t)} \frac{l - \tilde{l}}{ \mid l - \tilde{l} \mid^2} \; d \tilde{A}+\oint_{\bar{C}}\frac{\hat{t}}{2 \pi}\frac{\partial l}{\partial t} \cdot  \oint_{\bar{C}} \gamma_{\mathfrak{s}} (\tilde{s},t)\frac{l - \tilde{l}}{ \mid l - \tilde{l} \mid^2} \; d \tilde{s} + \frac{\hat{t}}{2 \pi}  \oint_{\bar{C}} \frac{\partial \gamma_{\mathfrak{s}}}{\partial t} (\tilde{s},t) \log \mid l - \tilde{l} \mid \; d \tilde{s} \right. \\ & \left. \hspace{0.5in} +\frac{\partial  \hat{t} }{ \partial t} \frac{ \omega}{2 \pi}\int_{D_v(t)} \log \mid l - \tilde{l} \mid \; d \tilde{A} +\frac{\partial  \hat{t} }{ \partial t}  \frac{1}{2 \pi}  \oint_{\bar{C}} \gamma_{\mathfrak{s}} (\tilde{s},t) \log \mid l - \tilde{l} \mid \; d \tilde{s} \right) \; ds \\
& \hspace{-0.5in}=\oint_{\bar{C}}\left( \frac{ \omega \hat{t}}{2 \pi}(v_{pi,n}+ v_{{si,n}}) \hat{n} \cdot  \int_{D_v(t)} \frac{l - \tilde{l}}{ \mid l - \tilde{l} \mid^2} \; d \tilde{A}+\oint_{\bar{C}}\frac{\hat{t}}{2 \pi}(v_{pi,n}+ v_{{si,n}}) \hat{n} \cdot  \oint_{\bar{C}} \gamma_{\mathfrak{s}} (\tilde{s},t)\frac{l - \tilde{l}}{ \mid l - \tilde{l} \mid^2} \; d \tilde{s} \right. \\
& + \frac{\hat{t}}{2 \pi}  \oint_{\bar{C}} \frac{\partial \gamma_{\mathfrak{s}}}{\partial t} (\tilde{s},t) \log \mid l - \tilde{l} \mid \; d \tilde{s} \\ & \hspace{1in}  +\frac{ \omega}{2 \pi}\frac{\partial}{\partial s} \bigg{[} (v_{pi,n}+ v_{{si,n}}) \hat{n} \bigg{]} \int_{D_v(t)} \log \mid l - \tilde{l} \mid \; d \tilde{A} \\ & \left. \hspace{1.5in}+ \frac{1}{2 \pi} \frac{\partial}{\partial s} \bigg{[} (v_{pi,n}+ v_{{si,n}}) \hat{n} \bigg{]} \oint_{\bar{C}} \gamma_{\mathfrak{s}} (\tilde{s},t) \log \mid l - \tilde{l} \mid \; d \tilde{s} \right) \; ds, 
\end{align*}
where $v_{pi,n}$ and $v_{si,n}$ denote, respectively, the terms within the curly brackets of equation (\ref{eq:cevol}).
 Every term in the last equation, except for the term on the second line, is determined at time $t$. This term contains the unknown $\partial \gamma_{\mathfrak{s}} / \partial t$. Introduce the following notation,
\[W \left(\frac{\partial \gamma_{\mathfrak{s}}}{\partial t},t \right):=\frac{d}{d t}  \oint_{\bar{C}} \Psi_i  \hat{t} \; ds.\]

Next, noting that for a sheet vortex, since $v_{po}=v_{pi}$ on the boundary, 
\[\gamma=\gamma_{\mathfrak{s}}+ \left(\left(E(l, \bar{C})-I \right) \cdot V_c(t) \right)^T \cdot \hat{t}, \]
where $I$ is the $2 \times 2$ identity matrix, a similar exercise shows that
\begin{align*}
&\frac{d}{d t}  \oint_{\bar{C}} \gamma l \; ds = \oint_{\bar{C}} \left(\frac{\partial  \gamma }{ \partial t} l +\gamma \frac{\partial  l }{ \partial t}  \right)\; ds, \\
&=\oint_{\bar{C}} \frac{\partial  \gamma_{\mathfrak{s}} }{ \partial t} l \; ds+ \frac{dV_c(t)^T}{dt} \cdot \oint_{\bar{C}} \left(E(l, \bar{C})-I \right)^T\cdot \hat{t} l\; ds + V_c(t) \cdot \oint_{\bar{C}} \mathbf{D}_t \left(E(l, \bar{C}) \right)^T\cdot \hat{t} l\; ds\\
&+V_c(t)^T \cdot \oint_{\bar{C}} \left(E(l, \bar{C})-I \right)^T\cdot \frac{\partial \hat{t}}{\partial t} l \; ds  + \oint_{\bar{C}} \gamma_{\mathfrak{s}}  \frac{\partial  l }{ \partial t}  \; ds +V_c(t)^T \cdot \oint_{\bar{C}}  \left(E(l, \bar{C})-I \right)^T\cdot \hat{t} \frac{\partial  l }{ \partial t}  \; ds, \\
&=\oint_{\bar{C}} \frac{\partial  \gamma_{\mathfrak{s}} }{ \partial t} l \; ds+ \frac{dV_c(t)^T}{dt} \cdot \oint_{\bar{C}} \left(E(l, \bar{C})-I \right)^T\cdot \hat{t} l\; ds + V_c(t) \cdot \oint_{\bar{C}} \mathbf{D}_t \left(E(l, \bar{C}) \right)^T\cdot \hat{t} l\; ds \\
& +V_c(t)^T \cdot \oint_{\bar{C}} \left(E(l,\bar{C})-I \right)^T\cdot \frac{\partial}{\partial s} \bigg{[} (v_{pi,n}+ v_{{si,n}}) \hat{n} \bigg{]} l \; ds + \oint_{\bar{C}} \gamma_{\mathfrak{s}}  (v_{pi,n}+ v_{{si,n}}
) \hat{n} \; ds \\
&\hspace{1in}+ V_c(t)^T \cdot \oint_{\bar{C}}  \left(E(l,\bar{C})-I \right)^T\cdot \hat{t} (v_{pi,n}+ v_{si,n}) \hat{n}  \; ds,
\end{align*}
The term $\mathbf{D}_t \left(E(l, \bar{C}) \right)^T$ in the above represents the time rate of change of the entries of $E$. Recalling that these entries are the first order spatial derivatives of the unit potentials $a$ and $b$, one can obtain a numerical estimate of these at time $t$ from $\bar{C}(s,t+\triangle)$ and applying the Laplace equation solver to numerically compute $a$ and $b$ at time $t+\triangle t$.

    The equation contains the unknowns $dV_c/dt$ and $\partial \gamma_{\mathfrak{s}}/\partial t$. Rewrite it as 
\begin{align*}
\frac{d}{d t}  \oint_{\bar{C}} \gamma l \; ds&=X \left(\frac{\partial \gamma_{\mathfrak{s}}}{\partial t},t \right)+\frac{dV_c(t)^T}{dt} \cdot \oint_{\bar{C}} \left(E(l)-I \right)^T\cdot \hat{t} l\; ds, \\
&=X \left(\frac{\partial \gamma_{\mathfrak{s}}}{\partial t},t \right)+ \left(\oint_{\bar{C}}  l^T \hat{t}^T \cdot \left(E(l)-I \right)\; ds \right) \cdot \frac{dV_c(t)}{dt}, 
\end{align*}
where in the last term elements of $\hat{t}^T \cdot \left(E(l)-I \right)$ are paired with $dV_c(t)/dt$, and $X$ is the vector denoting all the other terms on the right of the equation.

Next,
\begin{align*}
\frac{\partial v_p}{\partial t}&= -\frac{ \omega}{2 \pi}  \oint_{\bar{C}} \bigg{[}  \frac{\partial l}{\partial t} \cdot \frac{l- \tilde{l}}{ \mid l  - \tilde{l} \mid^2} \tilde{\hat{t}}+  \log \mid l  - \tilde{l} \mid   \frac{\partial \tilde{\hat{t}}}{\partial t}\bigg{]}\; d \tilde{s}, \\
&= -\frac{ \omega}{2 \pi}  \oint_{\bar{C}} \bigg{[} (v_{pi,n}+ v_{{si,n}}) \hat{n}  \cdot \frac{l- \tilde{l}}{ \mid l  - \tilde{l} \mid^2} \tilde{\hat{t}}+  \log \mid l  - \tilde{l} \mid \frac{\partial}{\partial \tilde{s}} \bigg{[} (\tilde{v}_{pi,\tilde{n}}+ \tilde{v}_{si, \tilde{n}}) \hat{\tilde{n}} \bigg{]}  \bigg{]}\; d \tilde{s},
\end{align*}
the`tilde' overhead in the last term again denoting that the $s$-parameter is replaced by $\tilde{s}$. And so $\partial v_p / \partial t$ is completely determined at time $t$.

Finally,
\begin{align*}
\frac{\partial}{\partial t} CPV&= - \frac{1}{2 \pi} \frac{\partial}{\partial t}\oint_{\bar{C}} \gamma_s(\tilde{s},t) \hat{k} \times  \frac{l  - \tilde{l}}{\mid l  - \tilde{l} \mid^2} \; d \tilde{s}, \\
&= - \frac{1}{2 \pi} \frac{\partial}{\partial t}\oint_{\bar{C}} \left[\frac{\partial \gamma_s(\tilde{s},t)}{\partial t}  \hat{k} \times  \frac{l  - \tilde{l}}{\mid l  - \tilde{l} \mid^2}\right. \\ 
& \left.  + \gamma_s(\tilde{s},t) \bigg{\{} (v_{pi,n}+ v_{{si,n}}) \hat{n} \cdot \nabla \left(\frac{l  - \tilde{l}}{\mid l  - \tilde{l} \mid^2} \cdot \hat{i} \right) \hat{i}+(v_{pi,n}+ v_{{si,n}}) \hat{n} \cdot \nabla \left(\frac{l  - \tilde{l}}{\mid l  - \tilde{l} \mid^2} \cdot \hat{j} \right) \hat{j}\bigg{\}}  \right]\; d \tilde{s}
\end{align*}and this equation also contains $\partial \gamma_{\mathfrak{s}}/\partial t$.

\end{document}